\documentclass{ws-procs10x7}
\def\rightnote{}
\def\cropmarks{}

\begin{document}

\title{Evidence for $B_s$ production at the $\Upsilon(5S)$ from CLEO and properties of the $\Upsilon(4S)$
from BaBar}

\author{J. C. Wang}

\address{Department of Physics, Syracuse University, Syracuse, NY 13244, USA\\
E-mail: jwang@physics.syr.edu}

\twocolumn[\maketitle\abstract{
BaBar experiment scan around the $\Upsilon(4S)$ resonance and measure its mass and full width.
They also measure ${\cal B}(\Upsilon(4S)\to B^0 \bar{B}^0) = 0.486\pm 0.010\pm 0.009$
from 81.7 fb$^{-1} \Upsilon(4S)$ data.
CLEO collaboration took about 0.42 fb$^{-1} \Upsilon(5S)$ data. 
They search for $B_s$ in both inclusive and exclusive modes
and find evidence for $B_s$ production at the $\Upsilon(5S)$ and
${\cal B}(\Upsilon(5S)\to B_s^{(*)}\bar{B}_s^{(*)}) = (21\pm 3\pm 9)\%$.}]

\section{Introduction}
The $\Upsilon(4S)$ is the lowest $b\bar{b}$ resonance above $B\bar{B}$ threshold.
Its mass and total width had been measured in scans of the total $e^+e^-$ cross-section
at center of mass energy around 10.58 
GeV.~\cite{Besson:1984bd,Lovelock:1985nb,Bebek:1987bp,Albrecht:1994et}
The $\Upsilon(4S)$ decays into $B^+B^-$ and $B^0\bar{B}^0$ modes allowing these particles
to be carefully studied.
Many $B$ branching fractions had been measured from $\Upsilon(4S)$ data.
Most of them, however, based on the assumption of equal production rates of the charged and 
neutral $B\bar{B}$ pairs.
Previous measurements are consistent with this 
assumption.~\cite{Alexander:2000tb,Aubert:2001xs,Athar:2002mr,Aubert:2004ur}
More precise measurement may result in renormalization of $B$ decay branching fractions.

The $\Upsilon(5S)$ was discovered by measuring the total hadronic cross-section 
above $\Upsilon(4S)$ as a function of energy at CESR.~\cite{Besson:1984bd,Lovelock:1985nb}
It is massive enough to produce $B_s^{(*)}\bar{B}_s^{(*)}$ pairs.
With limited data samples, experiments failed to clearly show the level of 
$B_s$ production at the $\Upsilon(5S)$.

In this paper I summarize recent studies from BaBar and CLEO on these issues.
The results on ${\cal B}(\Upsilon(4S)\to B^0\bar{B}^0)$ from BaBar and $B_s$ from
CLEO are preliminary.

\section{Evidence of $B_s$ in $\Upsilon(5S)$ at CLEO}

The $\Upsilon(5S)$ was discovered at CESR.~\cite{Besson:1984bd,Lovelock:1985nb}
Its mass was measured to be $(10.865 \pm 0.008)$ GeV.
It can decay into $B^{(*)}\bar{B}^{(*)}(\pi)$ modes, more channels than the $\Upsilon(4S)$
due to its heavier mass.
It is massive enough even to produce $B_s^{(*)}\bar{B}_s^{(*)}$ pairs.
Potential models predict about 1/3 of $\Upsilon(5S)$ produces $B_s^{(*)}\bar{B}_s^{(*)}$ pair.~\cite{bs_models}
The $B_s^*\bar{B}_s^*$ mode is the largest.
The experiments, however, failed to reveal if $B_s$ mesons were produced in about
0.1 fb$^{-1}$ data.

The CLEO III detector has recently recorded 0.42 fb$^{-1}$ of $e^+e^-$ annihilation data at 
the $\Upsilon(5S)$ resonance. Using this data sample they search for evidence of $B_s$ in both
inclusive and exclusive modes.~\cite{cleo_bs}

Most of $D_s$ production in $B_s$ decay is analogous to $D$'s in $B$ decay.
CLEO estimates ${\cal B}(\bar{B}_s \rightarrow D_s X) = (92\pm 11)\%$, whereas 
the measurement of ${\cal B}(\bar{B} \rightarrow D_s X) = (10.5\pm 2.6)\%$, which is 
the average of $B^+$ and $B^0$.
The distinct $D_s$ production rates can be used to unfold the production rate of $B_s$ in
$\Upsilon(5S)$ decays.

\begin{figure}
\begin{center}
\includegraphics[width=.32\textwidth]{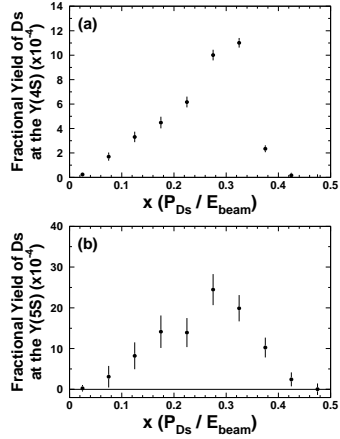}
\end{center}
\caption{The $D_s$ fractional yields vs x from (a)the $\Upsilon(4S)$ and (b)the $\Upsilon(5S)$ decays by CLEO,
 where the continuum contribution is subtracted.}
\label{fig:ds}
\end{figure}

CLEO reconstructs $D_s$ in the $D_s\to \phi \pi^+, \phi\to K^+K^-$ mode
from $\Upsilon(5S)$, $\Upsilon(4S)$ and continuum data.
The reconstruction efficiency is about 30\%. 
The $D_s$ yields as a function of x equal to the $D_s$ momentum divided by the beam energy
for $\Upsilon(4S)$ and $\Upsilon(5S)$ data are shown in Fig.~\ref{fig:ds}.
The contribution from $e^+e^-\to q\bar{q}$ events is subtracted,
the reconstruction efficiency is applied, but there is no correction for $D_s$ branching ratios.
The production of $D_s$ from $\Upsilon(5S)$ is significantly larger than that from $\Upsilon(4S)$.

Using ${\cal B}(D_s\to \phi \pi^+) = (3.6\pm 0.9)\%$, CLEO finds
\begin{eqnarray}
   {\cal B}(\Upsilon(4S) \to D_s X) & = (22.3 \pm 0.7 \pm 5.7)\%, \nonumber \\
   {\cal B}(\Upsilon(5S) \to D_s X) & = (55.0 \pm 5.2 \pm 17.8)\%, \nonumber
\end{eqnarray}
where the systematic error is dominated by $D_s$ decay branching ratio.
From these numbers CLEO finds
$${\cal B}(\Upsilon(5S) \to B_s^{(*)}\bar{B}_s^{(*)}) = (21 \pm 3 \pm 9)\%.$$
This is the first evidence of $B_s$ production at $\Upsilon(5S)$.
The rate agrees with theoretical expectations, which have a large range.

\begin{figure}
\begin{center}
\includegraphics[width=.3\textwidth]{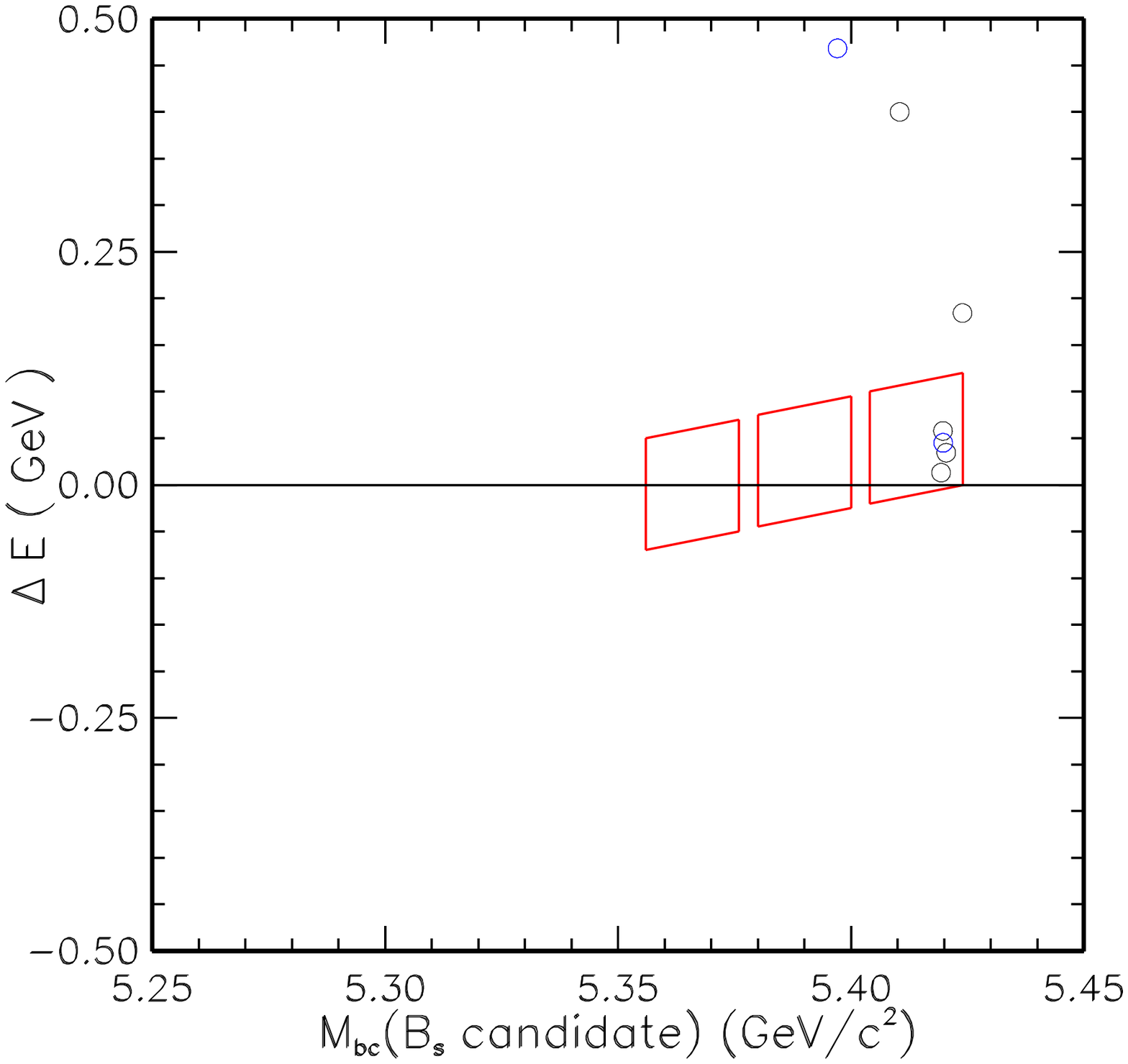}
\hspace{-0.01cm}
\includegraphics[width=.3\textwidth]{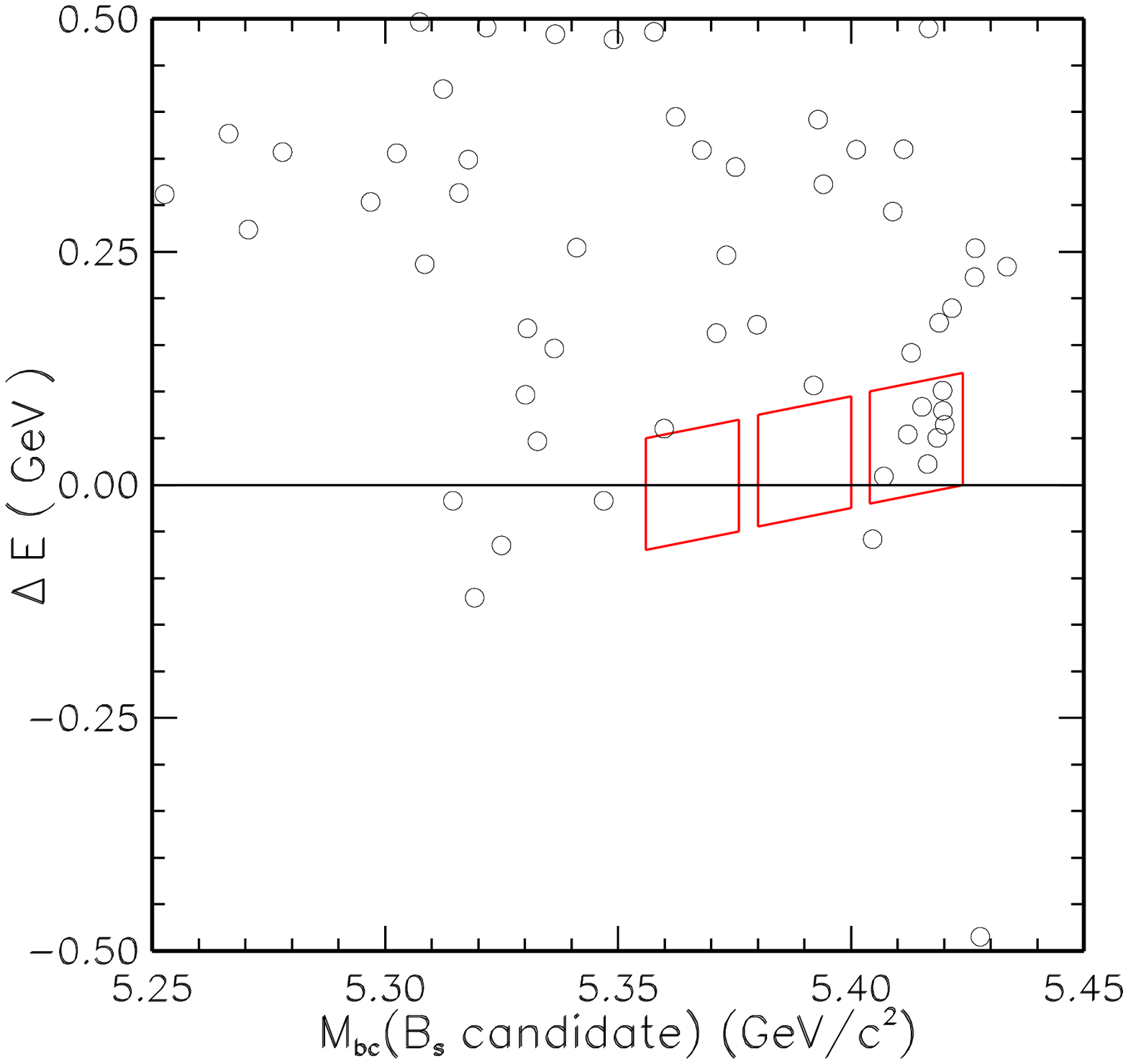}
\end{center}
\caption{The $M_{bc}$ vs $\Delta E$ distributions for (top) $\bar{B}_s\to J/\psi\ \phi/\eta'/\eta$ 
 and (bottom) $\bar{B}_s\to D_s^{(*)}\ \pi^-/\rho^-$ modes. 
The three signal boxes from left to right correspond to 
$\Upsilon(5S)\to B_s\bar{B}_s, B_s\bar{B}_s^*, B_s^*\bar{B}_s^*$ channels. }
\label{fig:bs}
\end{figure}

CLEO also looks for $B_s$ in two groups of exclusive modes: $\bar{B}_s\to J/\psi\ \phi/\eta'/\eta$
and  $\bar{B}_s\to D_s^{(*)}\ \pi^-/\rho^-$. The $M_{bc}$ vs $\Delta E$ plots
are shown in Fig.~\ref{fig:bs}, where the beam energy constraint mass and energy difference are
defined as
\begin{eqnarray}
M_{bc} & = \sqrt{E_{beam}^2 - P_{candidate}^2},\nonumber \\
\Delta E & = E_{beam} - E_{candidate}.
\end{eqnarray}
In the signal boxes 2 and 8 candidates for the two groups respectively are 
found.

The $B_s$ from $\Upsilon(5S)$ decays can be produced via three different channels:
$\Upsilon(5S)\to B_s\bar{B}_s, B_s\bar{B}_s^*, B_s^*\bar{B}_s^*$,
and one expects that ${\cal B}(B_s^*\to B_s\gamma) \sim 100\%$.
The energy of $B_s$ candidates produced through these three modes are 
not the same due to available kinetic energy and Lorentz boost,
resulting in 3 distinct signal regions as indicated in the plot.
The rightmost box where the candidates are found corresponds to 
the $B_s^*\bar{B}_s^*$ mode.
The large signal of  $B_s^*\bar{B}_s^*$ with respect to the other modes is
consistent with theoretical expectation.

\section{Measurement of $\Upsilon(4S)$ parameters}
The $\Upsilon(4S)$ is the lowest $b\bar{b}$ state above open bottom threshold.
The full width of $\Upsilon(4S)$, $\Gamma_{tot}$ is thus much larger than that of lower
$\Upsilon$ states, which allows direct measurement of its value at $e^+e^-$ collider.
The mass, total width and $e^+e^-$ partial width $\Gamma_{ee}$
had been previously measured by CLEO, CUSB and 
ARGUS.~\cite{Besson:1984bd,Lovelock:1985nb,Bebek:1987bp,Albrecht:1994et}
The values have relatively large uncertainty.
Different measurements show substantial variation. 
Improved measurements are necessary.

The BaBar detector is designed to operate at the SLAC PEP-II asymmetric-energy B factory.
The experiment scanned the $e^+e^-$ system at center of mass energy $\sqrt{s}$ around
the mass of $\Upsilon(4S)$, 10.58 GeV.~\cite{babar_y4s}


The $\Upsilon(4S)$ resonance parameters can be determined from the fit of visible
hadronic cross-section distribution to a so called line-shape function.
To the first order, BaBar uses relativistic Breit-Wigner function for the production 
cross section of $e^+e^-\to\Upsilon(4S)\to B\bar{B}$
\begin{equation}
 \sigma_0(s) = 12\pi\frac{\Gamma_{ee}^0 \Gamma_{tot}(s)}{(s-M^2)^2+M^2\Gamma_{tot}^2(s)}.
\label{eq:bw}
\end{equation}
The electric partial width $\Gamma_{ee}^0$ is taken as constant,
and the total width $\Gamma_{tot}(s)$  is energy dependent.
The function is further modified by radiative corrections calculable numerically, 
and the beam energy spread.
BaBar did one scan around $\Upsilon(3S)$ to determine the energy spread as well as energy calibration.
The visible hadronic cross-section also includes contributions from
$e^+e^-\to q\bar{q}$ continuum events and other processes 
which are not totally eliminated but suppressed.
This is modeled in the fit.
The integrated luminosity is measured using $e^+e^-\to\mu^+\mu^-$ process.

BaBar fit three cross section distributions simultaneously. 
The parameters of $\Upsilon(4S)$ are measured to be:
\begin{eqnarray}
  \Gamma_{tot}& = & (20.7 \pm 1.6 \pm 2.5)\ {\rm MeV},\nonumber \\
  \Gamma_{ee} & = & (0.321 \pm 0.017 \pm 0.029)\ {\rm keV},\nonumber\\
  M           & = & (10579.3 \pm 0.4 \pm 1.2)\ {\rm MeV}/c^2,\nonumber
\end{eqnarray}
where the uncertainty of energy spread, peak cross section, long term drift of energy scale, model
uncertainty and other sources are accounted for in the systematic errors.

\section{Measurement of ${\cal B}(\Upsilon(4S)\to B^0\bar{B^0})$}

The $\Upsilon(4S)$ decays into $B^+B^-$ and $B^0\bar{B^0}$ modes.
It is the most suitable environment to study $B$ physics. 
Many $B$ branching fractions had been measured from $\Upsilon(4S)$ data.
Most of the measurements, however, based on the assumption of equal production
rates of the charged and neutral $B\bar{B}$ pairs.
Theoretic models predict that the ratio of the charged pair production over neutral one
ranges from 1.03 to 1.25.~\cite{eichten}
Previous measurements are consistent to 1 within 
error.~\cite{Alexander:2000tb,Aubert:2001xs,Athar:2002mr,Aubert:2004ur}
A non-unit value of the ratio results in renormalization of the $B$ decay branching
fractions and contributes to our understanding of isospin violation in $B$ decays.

With a data sample of about 80 fb$^{-1}$ collected at $\Upsilon(4S)$ BaBar measured
${\cal B}(\Upsilon(4S)\to B^0\bar{B^0})$.~\cite{babar_f00}
The neutral mode is tagged with $\bar{B}^0\to D^{*+}l^-\nu$ decay.
The sample of events in which at least one  $\bar{B}^0\to D^{*+}l^-\nu$ candidate is found
is labeled as ``single-tag sample'', $N_s$.
The subset of ``single-tag sample'' where two candidates are found on both $B^0$ 
and $\bar{B}^0$ sides is labeled as the ``double-tag sample'', $N_d$.
We have
\begin{eqnarray}
  N_s & = & 2N_{B\bar{B}} f_{00} \epsilon_s {\cal B}(\bar{B}^0\to D^{*+}l^-\nu), \nonumber \\
  N_d & = &  N_{B\bar{B}} f_{00} \epsilon_d [{\cal B}(\bar{B}^0\to D^{*+}l^-\nu)]^2,
\end{eqnarray}
where total number of $B\bar{B}$ events $N_{B\bar{B}} = (88726\pm 23)\times10^3$,
$f_{00}={\cal B}(\Upsilon(4S)\to B^0\bar{B^0})$,
and $\epsilon_s$ and $\epsilon_d$ are the corresponding reconstruction efficiencies.
The double-tag reconstruction efficiency $\epsilon_d = \epsilon_s^2$ 
because the efficiencies are not correlated.
The ratio $f_{00}$ is thus given by
\begin{equation}
  f_{00} = \frac{N_s^2}{4N_d N_{B\bar{B}}}.
\end{equation}

\begin{figure}
\begin{center}
\includegraphics[width=.28\textwidth]{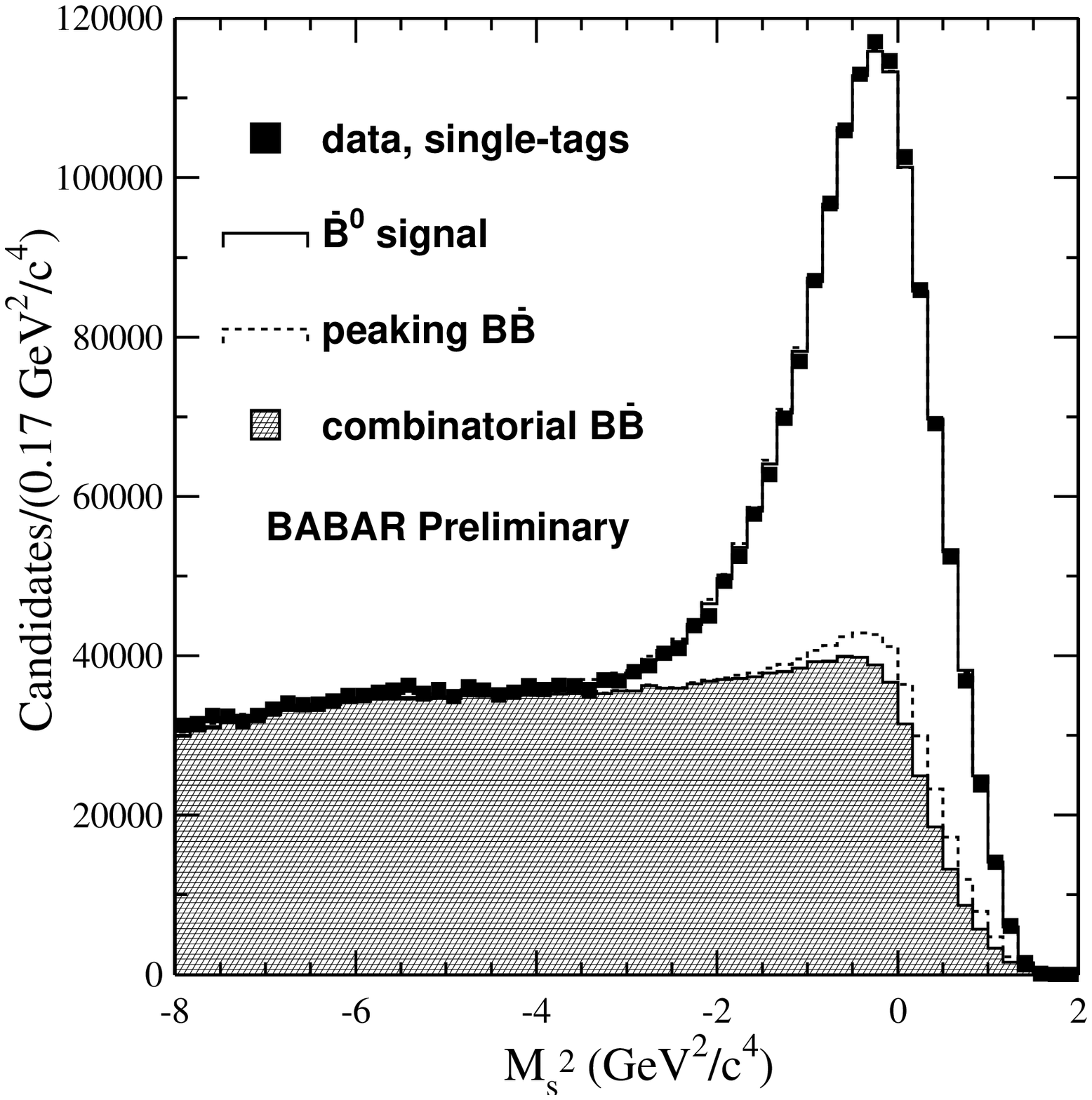}
\hspace{-0.01cm}
\includegraphics[width=.28\textwidth]{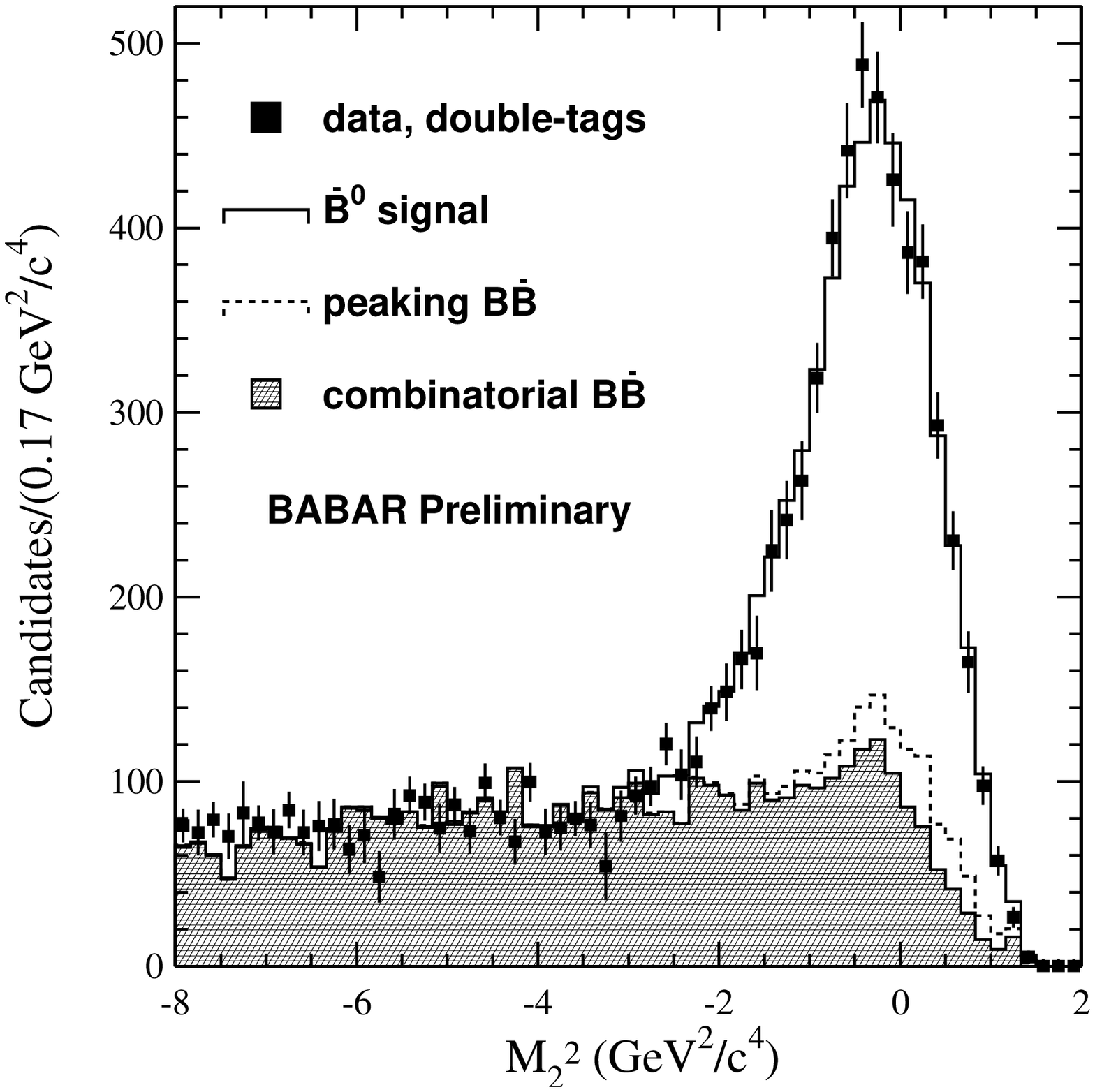}
\end{center}
\caption{The ${\cal M}^2$ distributions of single-tag (top) and double-tag (bottom) samples by BaBar.
Lines, and hatched area are fit to PDFs of various sources.}
\label{fig:f00}
\end{figure}

The measurement uses partial reconstruction of $\bar{B}^0\to D^{*+}l^-\nu$, where
only the lepton and the slow $\pi^+$ from $D^{*+}\to D^0\pi^+$ decay are observed.
This technique was first proposed by ARGUS~\cite{Albrecht:1993gr} and has been 
used in the CLEO measurement.~\cite{Athar:2002mr}
As there is very little kinematic energy released in $D^{*+}$ decay, momenta of 
$D^0$ and $\pi^+$ are correlated in $\Upsilon(4S)$ rest frame.
Thus $D^{*+}$ momentum can be parameterized with the $\pi^+$ momentum.
The neutrino invariant mass squared is calculated as:
\begin{equation}
  {\cal M}^2 \equiv (E_{beam}-E_{D^*}-E_l)^2 - (\vec{P}_{D^*}+\vec{P}_l)^2.
\end{equation}
The ${\cal M}^2$ distributions for single and double tag samples are shown in Fig.~\ref{fig:f00},
where contribution from $e^+e^-\to q\bar{q}$ is subtracted.

To determine $N_s$ and $N_d$, binned $\chi^2$ fits are performed to the two histograms.
The probability density functions(PDF) of signal and backgrounds are determined from MC simulation.
The number of signals are $N_s = 786300\pm 2000$ and $N_d=3560\pm80$.
The neutral branching rate, $f_{00} = 0.486 \pm 0.010 \pm 0.09$, 
is still consistent with equal production rates of the charged and neutral pairs.

\section*{Acknowledgments}
Many BaBar and CLEO colleagues helped in preparing this paper. 
The author would like to thank them, especially Professors S.~Stone, and I.~Shipsey, Doctors
R.~Godang, V.~Pavlunin and R.~Sia for useful discussions and comments.

\end{document}